\documentclass[twocolumn,aps,showpacs,prb,tightenlines,amsmath,amssymb]{revtex4}
\usepackage{amssymb}
\usepackage{amsmath}
\usepackage{colordvi}
\usepackage{graphicx}
\begin{document}
\title{Spin orbit coupling in bulk ZnO and GaN}
\author{J. Y. Fu}
\affiliation{Hefei National Laboratory for Physical Sciences at
Microscale, University of Science and Technology of China, Hefei,
Anhui, 230026, China}
\affiliation{Department of Physics,
University of Science and Technology of China, Hefei,
Anhui, 230026, China}
\author{M. W. Wu}
\thanks{Author to  whom correspondence should be addressed}
\email{mwwu@ustc.edu.cn.}
\affiliation{Hefei National Laboratory for Physical Sciences at
Microscale,
University of Science and Technology of China, Hefei,
Anhui, 230026, China}
\affiliation{Department of Physics,
University of Science and Technology of China, Hefei,
Anhui, 230026, China}
\altaffiliation{Mailing address.}
\date{\today}
\begin{abstract}
Using group theory and Kane-like $\mathbf{k\cdot p}$ model
together with the L\"owdining partition method, we
derive the  expressions of spin-orbit
coupling of electrons and holes,
including the linear-$k$ Rashba term
due to the intrinsic  structure inversion asymmetry
and the cubic-$k$ Dresselhaus
term due to the bulk inversion asymmetry in wurtzite semiconductors.
The coefficients of the electron and hole Dresselhaus terms
of ZnO and GaN in wurtzite structure and GaN in zinc-blende structure
are calculated using the nearest-neighbor $sp^3$  and $sp^3s^\ast$
tight-binding  models separately.
\end{abstract}
\maketitle

\section{Introduction}
The wide gap semiconductors ZnO, GaN and their alloys with
wurtzite or zinc-blende structures have
provoked a lot of interest in the last few years,
largely due to the good optical properties as well as great potential in
optoelectronics.\cite{Bagnall,Jain,Vurgaftman,rev1}
Very recently, some attention has also been given to the
spin properties of these semiconductors.\cite{Shen1,Shen2,Yan, Ghosh,Liu}
This is partly because of the prediction
that ZnO can become
ferromagnetic with a Curie temperature above room temperature if doped
with manganese.\cite{Dietl} Ferromagnetism and magnetoresistance in
Co-ZnO magnetic semiconductors have been also reported.\cite{Yan}
Moreover, long spin relaxation time observed in these
materials is another promising property for possible
applications in spintronic devices. Ghosh {\em et al.}\cite{Ghosh}
have investigated the
electron spin properties in $n$-type ZnO structures
and discovered the electron spin relaxation time varying from 20\ ns to 190\
ps when the temperature increases from 10 to 280\ K. Room
temperature electron spin relaxation as long as 25\ ns has also been
measured by electron paramagnetic resonance spectroscopy in
colloidal $n$-doped ZnO quantum dots.\cite{Liu}
In GaN, Beschoten {\em et al.}\cite{Beschoten}
reported  the  electron spin lifetime of 20\ ns
from $T=5$\ K to room temperature.
Recently, hole spin
relaxation time about 350\ ps at 1.7\ K in  ZnO epilayer has also been
reported.\cite{Lagarde}

Spin-orbit coupling (SOC) is the key issue of semiconductor
spintronics.\cite{Wolf,Zutic} Most of the proposed schemes of electrical
generation, manipulation and detection of electron or hole spins rely
on the SOC. A thorough understanding of the SOC is therefore very important.
In contrast to the zinc-blende  semiconductors such
as GaAs, the existence of hexagonal $c$ axis in wurtzite semiconductors
leads to an added intrinsic wurtzite  structure inversion asymmetry
(WSIA) in addition to the bulk inversion asymmetry (BIA).\cite{Bir,Weber}
Therefore, the  electron spin splittings include both the
Dresselhaus effect (cubic in $k$) and Rashba effect (linear
in $k$).\cite{Rashba,Bychkov} The Rashba effect has been
vigorously discussed using group theory and $\mathbf{k\cdot p}$
arguments by Lew Yan Voon {\em et al.}.\cite{Voon} It was pointed out
that the Rashba SOC coefficient is very
small (about 1.1\ meV$\cdot$\AA) in ZnO.\cite{Voon} Majewski and
Vogl reported a value of 9\ meV$\cdot$\AA\ of the Rashba coefficient
in GaN  based on the first principle calculation.\cite{Majewski}
Magneto-transport measurements in GaN
heterostructure gave the Rashba coefficient ranging from
5.5 to 10.01\ meV$\cdot$\AA .\cite{Kurdak,Thillosen,Schmult,Belyaev}
For electron Dresselhaus effect in wurtzite semiconductors, Wang {\em et
al.}  gave the form of electron Dresselhaus SOC
$H_{so}=\gamma_e(bk_z^2-k_{\|}^2)(\sigma_xk_y-\sigma_yk_x)$ recently
with $\gamma_e$ and $b$ the SOC parameters.
By fitting a spin degenerate surface near the $\Gamma$
point, they obtained $b=4.028$ and $\gamma_e\sim
0.74$\ meV$\cdot${\AA}$^3$ for AlN.\cite{Wang}
However, up to now, there is no report on the Dresselhaus coefficients
in ZnO and GaN.  Additionally, there is no investigation on the
SOC in valence bands.

In this work, by using group theory and $\mathbf{k\cdot p}$
method, we first construct the
$8\times8$ Kane model for wurzite structure,
including the $s$-$p_z$ mixing of the lowest
conduction band as well as
the contributions from the remote bands. Then we derive the forms
of the SOC for both electron and hole by L\"owdining partition
method.\cite{Lowdin}
The electron and hole Dresselhaus SOC
coefficients in wurtzite ZnO and GaN
are then investigated in detail by
using the $sp^3$ nearest-neighbor tight-binding
(TB) model, which was first put forward  for wurtzite semiconductors
by Kobayashi {\em et al.},\cite{Kobayashi} however, without the
SOC effect. We incorporate the SOC following the approach by Chadi.\cite{Chadi}

This paper is organized as follows: In Sec.\ II, we present the model.
We start with the  Kane-like
$\mathbf{k\cdot p}$ model for wurtzite semiconductors and
derive the expressions for the SOC for electron and hole
perturbatively up to third order in Sec.\ IIA.
In Sec.\ IIB, we briefly introduce the nearest-neighbor $sp^3$ TB model
for wurtzite semiconductors.
We present our main results in Sec.\ III and conclude  in  Sec.\ IV.

\section{Model and Hamiltonian}
In contrast to the form given by Chuang and
Chang,\cite{Chuang} in which the $s$-$p_z$ mixing of the lowest
conduction band and the contributions from the remote bands are
missing, in this section,
we first construct  the 8$\times$8 Kane-like $\mathbf{k\cdot p}$ model for
wurtzite semiconductors with the $s$-$p_z$ mixing properly included.
 The contributions from the remote bands are also
considered perturbatively.
Then we derive the expressions of SOC for both electron and hole.
Finally, we briefly introduce the $sp^3$ TB model for the wurtzite
semiconductors which we use for obtaining the SOC coefficients.

\subsection{Kane-like $\mathbf{k\cdot p}$ model}
The Schr\"odinger equation relating the periodic part
$u_{\nu\mathbf{k}}(\mathbf{r})$ of the Bloch
function and the energy near the band edge has the form\cite{Kane,Winkler}
\begin{eqnarray}
  \label{eq1}
\nonumber
  Hu_{\nu\mathbf{k}}(\mathbf{r})
&\approx&(H_0+\frac{\hbar^2k^2}{2m_0}+\frac{\hbar}{m_0}\mathbf{k\cdot p}+H_{so})u_{\nu\mathbf{k}}(\mathbf{r})\\
&=&E_n(\mathbf{k})u_{\nu\mathbf{k}}(\mathbf{r}),
\end{eqnarray}
where
  \begin{eqnarray}
 \label{eq2}
H_0&=&\frac{p^2}{2m_0}+V(\mathbf{r}),\\
H_{so}&=&\frac{\hbar}{4m_0^2c^2}\nabla
    V\times\mathbf{p}\cdot\mbox{\boldmath$\sigma$\unboldmath}\nonumber\\
&=&H_{sx}\sigma_x+H_{sy}\sigma_y+H_{sz}\sigma_z.
  \label{eq3}
\end{eqnarray}
Here $V({\bf r})$ is the periodic potential, $H_{so}$ accounts for the
spin-orbit interaction, and $\sigma_i$ with $i=x,y,z$ are the Pauli spin
matrices.

Following Chuang and Chang, the basis functions used near the zone
center read $|u_1\rangle = |iS\uparrow\rangle$,
$|u_2\rangle = \Big|-\frac{X+iY}{\sqrt{2}}\uparrow\rangle$,
$|u_3\rangle =\Big|\frac{X-iY}{\sqrt{2}}\uparrow\rangle$,
$|u_4\rangle = |Z\uparrow\rangle$,
$|u_5\rangle = |iS\downarrow\rangle$,
$|u_6\rangle = \Big|\frac{X-iY}{\sqrt{2}}\downarrow\rangle$,
$|u_7\rangle = \Big|-\frac{X+iY}{\sqrt{2}}\downarrow\rangle$, and
$|u_8\rangle = |Z\downarrow\rangle$.\cite{Chuang}
Here
$|S\rangle$, $|X\rangle$, $|Y\rangle$, and $|Z\rangle$ represent
the symmetry of the band edge states, and the arrows stand for the
spin orientation. The $z$ direction corresponds to the $c$ axis of the
wurtzite crystal. From the $C_{6v}$ symmetry analysis of
point group for the wurtzite structure, the $\mathbf{k\cdot p}$ Hamiltonian
in the basis of $|u_i\rangle$ with $i=1,\cdots,8$
can be written as
\begin{widetext}
\begin{eqnarray}
  \label{eq4}
\nonumber
 H_{8\times8} &=& \frac{\hbar^2k^2}{2m_0}
  +
 \setlength{\tabcolsep}{2pt}
 \renewcommand{\arraystretch}{2.0}
 \left( \begin{array}{cccccccc}
  E_c+ \atop
     \ \ \ \frac{\hbar^2k^2}{2m^\prime} & -\frac{P_2}{\sqrt{2}}k_+- \atop
      iB_{cv1}k_zk_+  &
      \frac{P_2}{\sqrt{2}}k_-
      &  P_1k_z- \atop
     \ \ \ iB_{cv2}k_{\|}^2 & 0  & 0
      & -\sqrt{2}i\Delta_{sz} & 0 \\
     -\frac{P_2}{\sqrt{2}}k_-+ \atop
      iB_{cv1}k_zk_- &  E_v+\Delta_1+ \atop
    \Delta_2+A^\prime k_z^2
      & 0 & \frac{iQ}{\sqrt{2}}k_-+ \atop
    \ \ \ B_{cv3}k_zk_-
      & 0 &
      0 & 0 & 0 \\
     \frac{P_2}{\sqrt{2}}k_+& 0 & E_v+\Delta_1- \atop
    \hspace{-0.6cm}\Delta_2
      & -\frac{iQ}{\sqrt{2}}k_+  &  \sqrt{2}i\Delta_{sz}& 0
    & 0 & \sqrt{2}\Delta_3 \\
     P_1k_z+ \atop
     \ \ \ iB_{cv2}k_{\|}^2 &-\frac{iQ}{\sqrt{2}}k_++ \atop
    \ \ \ B_{cv3}k_zk_+  & \frac{iQ}{\sqrt{2}}k_- & E_v+ \atop
    \ \ \ B^\prime k_{\|}^2
    & 0 & 0 & \sqrt{2}\Delta_3 & 0 \\
    \frac{P_2}{\sqrt{2}}k_+ & 0 & -\sqrt{2}i\Delta_{sz} & 0 &
    E_c+ \atop
     \ \ \ \frac{\hbar^2k^2}{2m^\prime}
    & \frac{P_2}{\sqrt{2}}k_-+ \atop
     iB_{cv1}k_zk_- & -\frac{P_2}{\sqrt{2}}k_+ &
     P_1k_z- \atop
     \ \ \ iB_{cv2}k_{\|}^2 \\
    0 & 0 & 0 & 0 & \frac{P_2}{\sqrt{2}}k_+- \atop
     iB_{cv1}k_zk_+ &
     E_v+\Delta_1+ \atop
    \Delta_2+A^\prime k_z^2
     & 0 &  -\frac{iQ}{\sqrt{2}}k_+- \atop
    \ \ \ B_{cv3}k_zk_+  \\
    \sqrt{2}i\Delta_{sz} & 0 & 0 & \sqrt{2}\Delta_3  &
     -\frac{P_2}{\sqrt{2}}k_-  &
      0 &  E_v+\Delta_1- \atop
    \hspace{-0.6cm}\Delta_2 & \frac{iQ}{\sqrt{2}}k_- \\
    0 & 0 & \sqrt{2}\Delta_3 & 0 & P_1k_z+ \atop
     \ \ \ iB_{cv2}k_{\|}^2 & \frac{iQ}{\sqrt{2}}k_-- \atop
    \ \ \ B_{cv3}k_zk_-
    & -\frac{iQ}{\sqrt{2}}k_+ &  E_v+ \atop
    \ \ \ B^\prime k_{\|}^2 \\
  \end{array} \right),  \\ 
\end{eqnarray}
\end{widetext} 
in which $k_{\pm}=k_x\pm ik_y$.
The energy parameters are defined by
 $\langle S|H_0| S\rangle = E_c$,
$\langle X|H_0| X\rangle=\langle Y|H_0| Y\rangle=E_v+\Delta_1$,
$\langle Z|H_0| Z\rangle = E_v$,
$\langle Y|H_{sz}| X\rangle = i\Delta_2$, and
$\langle Z|H_{sx}| Y\rangle =-\langle Z|H_{sy}| X\rangle = i\Delta_3$.
The interband momentum-matrix elements read
$P_1 = -\frac{i\hbar}{m_0}\langle S|p_z|Z\rangle$,
$P_2 = -\frac{i\hbar}{m_0}\langle S|p_x|X\rangle=-\frac{i\hbar}{m_0}\langle
S|p_y|Y\rangle$, and
$Q = -\frac{i\hbar}{m_0}\langle Z|p_x|X\rangle=-\frac{i\hbar}{m_0}\langle
Z|p_y|Y\rangle$. It is noted that the parameter $Q$ was omitted in Ref.\ \onlinecite{Chuang}.
The inclusion of $Q$ gives an improved description of the
  valence bands. It also determines the hole spin splitting.\cite{Dugdale}
The other parameters $m^\prime$, $B_{cv1}$, $B_{cv2}$,
$B_{cv3}$, $A^\prime$, and $B^\prime$ are
related with the contributions from the remote bands,
which 
were also omitted in Ref.\ \onlinecite{Chuang}.
$\Delta_{sz}$ originates from the $s$-$p_z$ mixing 
of the lowest conduction band and
has the same  definition as the energy parameter $\Delta_3$.
In addition,  $\Delta_{sz}$ is closely related to
the linear Rashba terms of both electron and hole.\cite{Voon,Ikai}
This term was also omitted in Ref.\ \onlinecite{Chuang}.
As for the energy parameters $\Delta_1$, $\Delta_2$ and $\Delta_3$,
one can relate them to
crystal-field split energy $\Delta_{cr}$ and the spin-orbit split-off
energy $\Delta_{so}$ by
\begin{equation}
\label{eq5}
\Delta_1=\Delta_{cr}, \ \ \ \ \Delta_2=\Delta_3=\frac{1}{3}\Delta_{so}.
\end{equation}

Using L\"owdining partition method,\cite{Lowdin}
one can derive the SOC terms for electron and
hole, which act as  $k$-dependent effective magnetic fields.
For the states near $\Gamma$ point, the electron SOC
can be described
by $[\alpha_e\mathbf{\Omega}_e^R(\mathbf{k})+
\gamma_e\mathbf{\Omega}_e^D(\mathbf{k})]
\cdot\mbox{\boldmath$\sigma$\unboldmath}$
with
\begin{eqnarray}
\label{eq6}
&&\alpha_e\mathbf{\Omega}_e^R(\mathbf{k})=\alpha_e(k_y,-k_x,0),\\
\label{eq7}
&&\gamma_e\mathbf{\Omega}_e^D(\mathbf{k})=\gamma_e(bk_z^2-k_{\|}^2)(k_y,-k_x,0).
\end{eqnarray}
Here $\alpha_e\mathbf{\Omega}_e^R(\mathbf{k})$ and
 $\gamma_e\mathbf{\Omega}_e^D(\mathbf{k})$ are
the  Rashba  and  Dresselhaus terms,
respectively. $\alpha_e$ and $\gamma_e$ are the corresponding SOC
coefficients. $b$ is a parameter that can be determined by the TB calculation.
It should be mentioned that $\alpha_e$ originates from the $s$-$p_z$
mixing of the lowest conduction band, whereas $\gamma_e$ is
closely related with $\mathbf{k\cdot p}$ interaction with
the remote bands having $\Gamma_6$ symmetry.

As we  know, for zinc-blende structure, both the heavy hole (HH) and
light hole (LH) belong to the four dimensional
$\Gamma_{8v}$ group presentation  whereas the spin split-off hole (SOH) has the
$\Gamma_{7v}$ symmetry.\cite{Winkler,Pfeffer} However,
for wurtzite structure, the combination of crystal-field and SOC
energies lead to a three-edge structure at the
top of the valence band. Two of these three edges are of $\Gamma_{7v}$ symmetry
and the remaining one is of $\Gamma_{9v}$ symmetry.
The symmetry of the valence bands are, in the order of
decreasing energy, $\Gamma_{9v}$ (HH), $\Gamma_{7v}$ (LH), and
$\Gamma_{7^\prime v}$
(crystal-field split-off hole) for
GaN\cite{Zunger,Kumagai} and
 $\Gamma_{7v}$ (HH), $\Gamma_{9v}$ (LH), and $\Gamma_{7^\prime v}$
 (crystal-field split-off hole)
 for ZnO.\cite{Voon,Xia}
This anomalous ordering in ZnO results from a negative spin-orbit
splitting.\cite{Rowe} Hereafter, in the description of the 
SOC of holes of ZnO and GaN, 
we use the subscript ``9'', ``7'', and ``$7^\prime$''
from the symmetry representation to label the three kinds of holes above.
For the lowest conduction band, both ZnO and GaN have  $\Gamma_{7c}$
symmetry and we use the symbol ``$e$'' to denote it.
The corresponding SOC terms are $[\alpha_i\mathbf{\Omega}_i^R(\mathbf{k})+
\gamma_i\mathbf{\Omega}_i^D(\mathbf{k})]\cdot\mbox{\boldmath$\sigma$\unboldmath}$,
which include both the linear Rashba term
$\alpha_i\mathbf{\Omega}_i^R(\mathbf{k})$,
and the cubic Dresselhaus term $\gamma_i\mathbf{\Omega}_i^D(\mathbf{k})$,
with $i=e,9,7,7^\prime$. Their expressions  are
given in Table\ \ref{table1}.

The Rashba coefficients $\alpha_i$ with $i=e,7,7^\prime$
are expressed  using the
$\mathbf{k\cdot p}$ parameters:
\begin{eqnarray}
\alpha_e&=&\frac{2P_2\Delta_{sz}}{E_g+\Delta_{so}},\\
\label{eq8}
\alpha_7&=&2(\frac{Q\Delta_{so}}{3\Delta_{cr}}-\frac{P_2\Delta_{sz}}{E_g+\Delta_{so}}),\\
\alpha_{7^\prime}&=&\frac{-2Q\Delta_{so}}{3\Delta_{cr}},
\end{eqnarray}
where $E_g$ is the energy gap between the bottom of the conduction
band and the top of the valence band.
As for $\Gamma_{9v}$ symmetry, there is no linear-$k$ splitting,
namely the Rashba coefficient
$\alpha_9=0$, as pointed out in previous literature.\cite{Casella,Mahan}
Moreover, it is noted for the expression of $\alpha_7$ [Eq.~(\ref{eq8})]
that the first term dominates due to
the large band gap in ZnO and GaN. Therefore, due to
the negative spin splitting energy in ZnO, both $\alpha_e$ and $\alpha_7$ are negative.
However, they are positive for GaN. The situation of $\alpha_{7^\prime}$ is reversed.

\begin{table*}[htb]
  \caption{Electron and hole SOC terms.
$\alpha_i\Omega_i^R (\mathbf{k})$
and $\gamma_i\Omega_i^D (\mathbf{k})$ represent the Rashba and
Dresselhaus terms with $i=e,9,7,7^\prime$ for wurtzite structure
 and  $i=e,8,7$ for zinc-blende structure.
 $\alpha_i$ and $\gamma_i$ are
the Rashba and Dresselhaus SOC coefficients.}
\begin{tabular}{lllllllllllllll}\hline\hline
1. Wurtzite structure & \mbox{\hspace{2cm}} $\alpha_i\Omega_i^R (\mathbf{k})$
&\mbox{\hspace{3.0cm}}  $\gamma_i\Omega_i^D (\mathbf{k})$ & \\ \hline
Conduction band ($\Gamma_{7c}$):
& \mbox{\hspace{1.5cm}}$\alpha_e(k_y,-k_x,0)$& \mbox{\hspace{1.5cm}}
 $\gamma_e(bk_z^2-k_{\|}^2)(k_y,-k_x,0)$ & \\\hline
\hspace{2.12cm}($\Gamma_{9v}$):
  &\mbox{\hspace{2.3cm}} 0 & \mbox{\hspace{1.5cm}}
$\gamma_9(k_y(k_y^2-3k_x^2),k_x(k_x^2-3k_y^2),0)$& \\
Valence bands ($\Gamma_{7v}$):
&\mbox{\hspace{1.5cm}} $\alpha_7(k_y,-k_x,0)$ &\mbox{\hspace{1.5cm}}
$\gamma_7(bk_z^2-k_{\|}^2)(k_y,-k_x,0)$ & \\
\hspace{2.12cm}($\Gamma_{7^\prime v}$):
& \mbox{\hspace{1.6cm}}$\alpha_{7^\prime}(k_y,-k_x,0)$ &\mbox{\hspace{1.5cm}}
$\gamma_{7^\prime}(bk_z^2-k_{\|}^2)(k_y,-k_x,0)$ &  \\
\hline\hline
\\
\hline\hline
2. Zinc-blende structure \cite{Winkler} & \mbox{\hspace{2cm}}
 $\alpha_i\Omega_i^R (\mathbf{k})$
&\mbox{\hspace{3.0cm}}  $\gamma_i\Omega_i^D (\mathbf{k})$&\\ \hline
Conduction band ($\Gamma_{6c}$):
&\mbox{\hspace{2.3cm}} 0 &\mbox{\hspace{0.6cm}}
$\gamma_e(k_x(k_y^2-k_z^2),k_y(k_z^2-k_x^2),k_z(k_x^2-k_y^2))$ \\\hline
Valence bands ($\Gamma_{8v}$):
 &\mbox{\hspace{2.3cm}} 0 &\mbox{\hspace{0.6cm}}
$\gamma_8(k_x(k_y^2-k_z^2),k_y(k_z^2-k_x^2),k_z(k_x^2-k_y^2))$\\
\hspace{2.12cm}($\Gamma_{7v}$):
&\mbox{\hspace{2.3cm}} 0 &\mbox{\hspace{0.6cm}}
$\gamma_7(k_x(k_y^2-k_z^2),k_y(k_z^2-k_x^2),k_z(k_x^2-k_y^2))$  \\
\hline\hline
\end{tabular}
\label{table1}
\end{table*}

As for the Dresselhaus effect, both the SOC coefficient $\gamma_i$
and the parameter $b$
can be determined by the TB calculation.
While it should be mentioned that $\gamma_i$ is closely related
with the remote bands with $\Gamma_6$ symmetry.
It is also noted that the electron Rashba and Dresselhaus terms
of ZnO and GaN have opposite signs. 
However, it is  not the case for  $\Gamma_{7v}$ and
$\Gamma_{7^\prime v}$ holes. For the
$\Gamma_{9v}$ holes, only the Dresselhaus term exists.

For the sake of comparison, SOC terms of zinc-blende
structures\cite{Winkler}
are also listed in Table\ \ref{table1}. 
For $\Gamma_{8v}$ valence band, the SOC
form reads $\gamma_8\mathbf{\Omega}_8^D\cdot
\mathbf{J}$. Here the small modifications
 from the $\mathbf{k\cdot p}$ interactions with the remote
$\Gamma_3$ state and the SOC within the conduction band
$\Gamma_{8c}$  are omitted.
The matrices $J_i$  with $i=x,y,z$ are angular momentum matrices
for angular momentum $J=3/2$.

\subsection{TB model}
To calculate electron and hole spin splittings, we use the $sp^3$ TB
model elaborated by Kobayashi {\em et al.},\cite{Kobayashi}
which has been proven to be an effective approach in band
structure calculations for wurtzite crystals.\cite{Kobayashi,Jenkins,Malkova}
In this model, the local point symmetry is approximately tetrahedral,
namley $T_d$ symmetry rather than $C_{3v}$ symmetry.
In such an approximation, on-site coupling
between $s$ and $p_z$  orbitals, and the small crystal field
splittings between $p_z$ orbital and the $p_{x,y}$ orbital are
both  neglected. Thus the model has nine
independent parameters: the four on-site matrix elements $E(s,a)$,
$E(p,a)$, $E(s,c)$, $E(p,c)$ (where $s$ and $p$ refer to the basis
states, and $a$ and $c$ refer to anion and cation), and five
transfer matrix elements $V(ss\sigma)$,
$V(sp\sigma)$, $V(ps\sigma)$, $V(pp\pi)$, and $V(pp\sigma)$ [where the
orientation of the $p$ orbitals is denoted by $\sigma$ and $\pi$, and
the first (second) index refers to the $s$ or $p$ state of
 anion (cation)].
For numerical calculations, we use the TB
parameters taken from Refs.\ \onlinecite{Kobayashi} and \onlinecite{Jenkins}.

The SOC is included as outlined by Chadi.\cite{Chadi}
We take the spin splitting parameters $\Delta_{\mbox{\tiny
Zn}}$=0.335\ eV, $\Delta_{\mbox{\tiny O}}$=0.0274\ eV,
$\Delta_{\mbox{\tiny Ga}}$=0.174\ eV, and
$\Delta_{\mbox{\tiny N}}$=0.009\ eV.\cite{Chadi,Shindo,Phillips} It
should be mentioned that only the cubic term, namely the Dresselhaus spin
splitting can be calculated by the TB
model under $T_d$ symmetry approximation.
Based on the zero spin splitting point calculated by the TB
  model, one can conveniently fit the parameter $b$ in
 the Dresselhaus SOC descriptions listed in Table\ \ref{table1}.
Moreover, the SOC coefficient $\gamma_i$ can also be obtained
from the calculated spin splitting energy.

\begin{figure}[bth]
  \centering
\includegraphics[height=5.5cm]{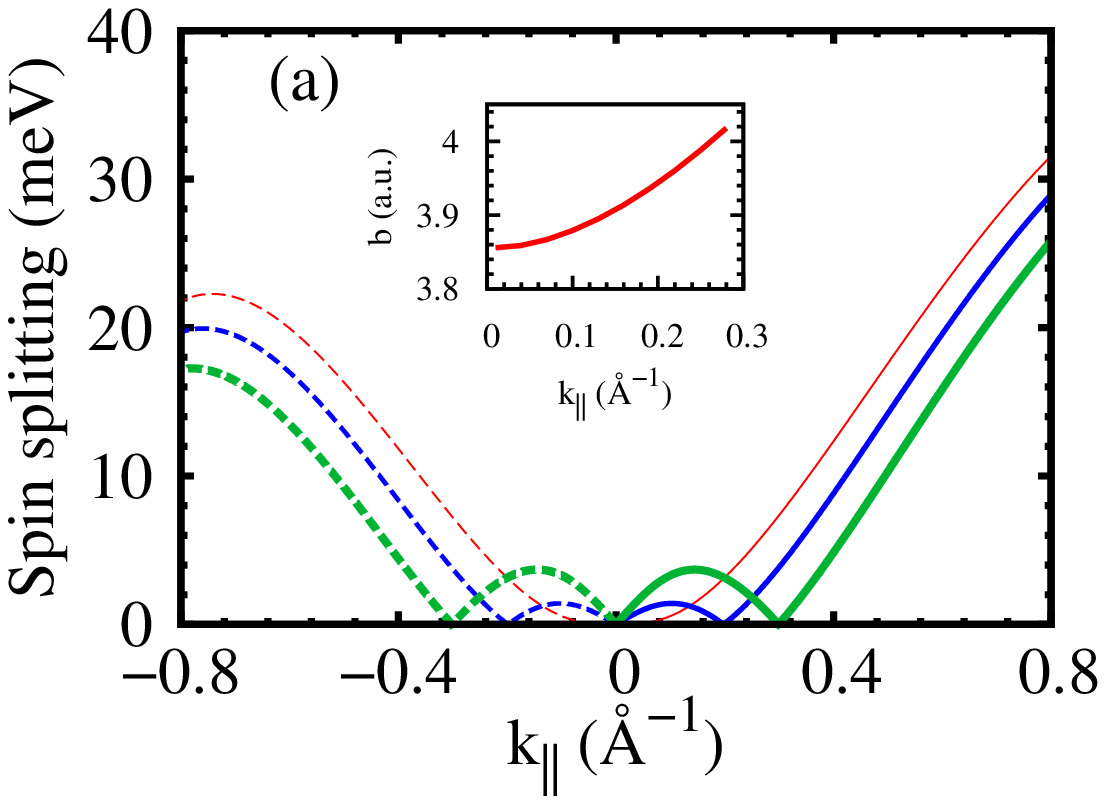}
  \includegraphics[height=5.5cm]{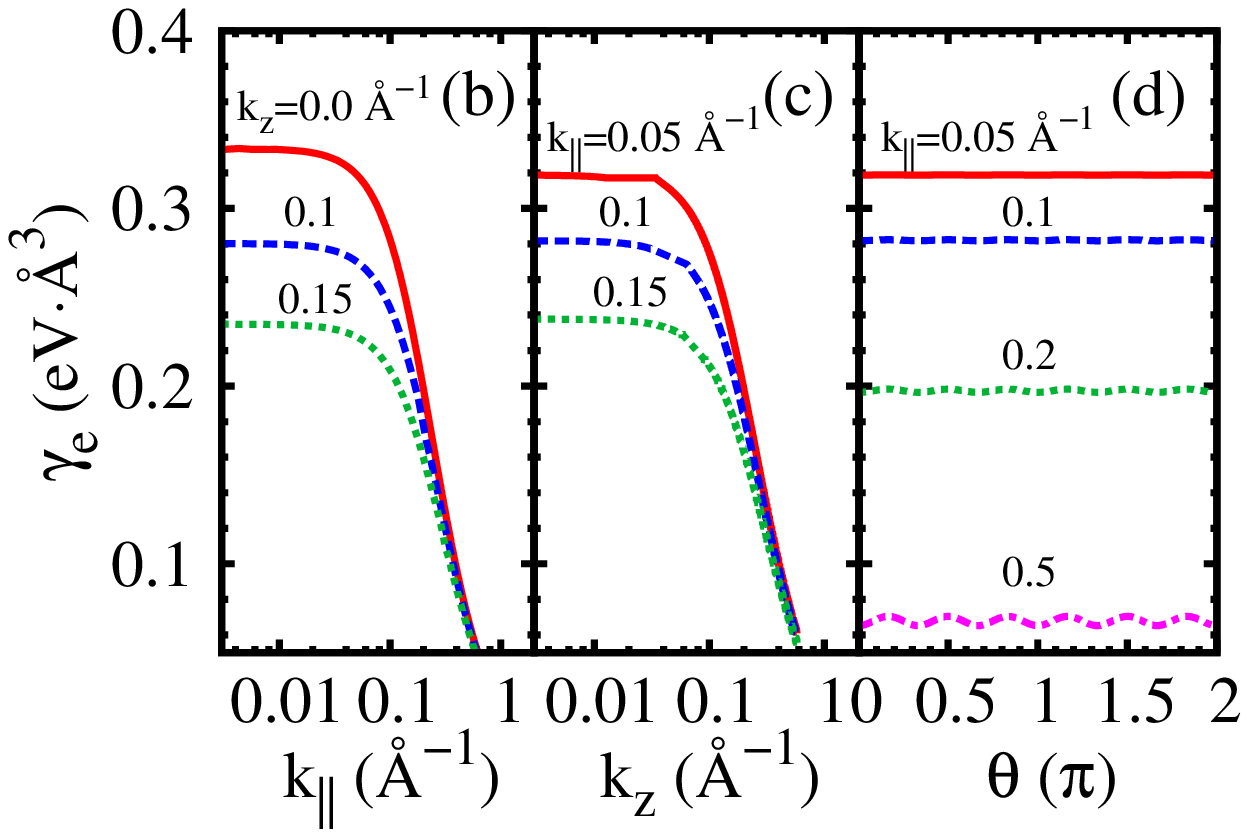}
 \caption{(Color Online) (a) Electron spin splitting of
ZnO against in-plane momentum along  $\Gamma$-$K$
(solid  curves) and $\Gamma$-$M$ (dashed curves)
    directions. Red (thin) curves: $k_z$=0; Blue curves:
$k_z$=0.1\ {\AA}$^{-1}$; Green (thick)
curves: $k_z$=0.15\ {\AA}$^{-1}$. Inset: The coefficient $b$ as function of
$k_\|$. (b) $\gamma_e$ {\it vs}. $k_{\|}$
along $\Gamma$-$K$ direction under different $k_z$.
    (c) $\gamma_e$ {\it vs}. $ k_z$
under different $k_\|$.
    (d) $\gamma_e$ {\em vs}. angle of the in-plane momentum
at different $k_\|$ in the plane of $k_z=0$.}
  \label{fig1}
\end{figure}

\section {Results}
Since the linear Rashba term has been  investigated in
Ref.\ \onlinecite{Voon}, in this section we mainly discuss the cubic
Dresselhaus terms of wurtzite ZnO and GaN, calculated from the nearest-neighbor
$sp^3$ TB model. In fact, the electron Dresselhaus term
becomes the dominant SOC compared with the linear
Rashba term  when the electron concentration is higher than
$10^{19}$\ cm$^{-3}$ for ZnO and $10^{20}$\ cm$^{-3}$ for GaN.
(Note in the literature,  the electron concentrations of ZnO and GaN
are in the range of $10^{17}$~cm$^{-3}$ to $10^{21}$~cm$^{-3}$.\cite{Izyumskaya,Ghosh,Belyaev}) In order to
compare with the zinc-blende structure, the cubic case of
GaN is also addressed.

\subsection{SOC in ZnO}

In Fig.\ \ref{fig1}(a), we show the cubic electron spin splitting energy
around $\Gamma$ point for
different momentum in the planes of $k_z$=0 (red thin curves), 0.1 (blue
curves), and 0.15\ \AA$^{-1}$ (green thick curves), respectively. In each
$k_z$ plane, the results along  $\Gamma$-$K$ (solid
curves) and $\Gamma$-$M$ (dashed curves) directions are shown
separately. In contrast to the spin splitting
of GaAs at $\Gamma$ point,\cite{Wu} one can see from the
figure that, for small momentum, the
splittings show isotropy. Moreover, in the case of $k_z$=0.1\ \AA$^{-1}$, there is
a zero splitting point at $k_{\|}$=0.198\ \AA$^{-1}$,
from which one can determine that $b$ in the Dresselhaus term is  3.910.
Furthermore,  we find from our calculation that there is a certain variation
of $b$  as
 shown in the inset of Fig.\ \ref{fig1}(a). From the
inset one notices  that $b$ increases monotonically  with
 $k_{\|}$. It also implies that for small momentum, $b$
keeps constant around 3.855.  Moreover,
 it is independent on the angle of the in-plane wave
vector $k_{\|}$, which can be
seen from the fact of the same zero splitting points between $\Gamma$-$K$ and
$\Gamma$-$M$ directions under the same  wave vector $k_z$.

Up till now, there has been no
report on the  value of $\gamma_e$. In this work, the value
of this coefficient is extracted from
\begin{equation}
\gamma_e(\mathbf{k})=\Delta E/(2|\Omega_e^D(\mathbf{k})|),
\end{equation}
with $\Delta E$ standing for
the corresponding Dresselhaus spin splitting energy. The results
for different momentum are shown in Fig.\ \ref{fig1}(b-d).
One can see from the figure that for small momentum, $\gamma_e$
is almost a constant
(about 0.33~eV$\cdot${\AA}$^3$) with its value being almost two orders
of magnitude  smaller
than the electron Dresselhaus coefficient 17.0\
eV$\cdot${\AA}$^3$ in GaAs, \cite{Wu}  thanks to the large
band gap ($E_g$ $\sim$ 3.3~eV at 300\ K) in ZnO.\cite{Morkoc}
When the momentum lies far away from the $\Gamma$ point, our calculation
indicates that the value of $\gamma_e$ decreases,
either with the increase of the in-plane wave vector $k_{\|}$
in Fig.\ \ref{fig1}(b),
or with the out-of-plane wave vector $k_z$
in Fig.\ \ref{fig1}(c).
This $k$-dependence of $\gamma_e$ is due to the correction of the
higher order SOC terms for large momentum.
Moreover, angle dependence of
$\gamma_e$ becomes remarkable for large momentum, which is also
because of the correction of higher order SOC terms.

\begin{figure}[bth]
  \centering
 \includegraphics[height=5.5cm]{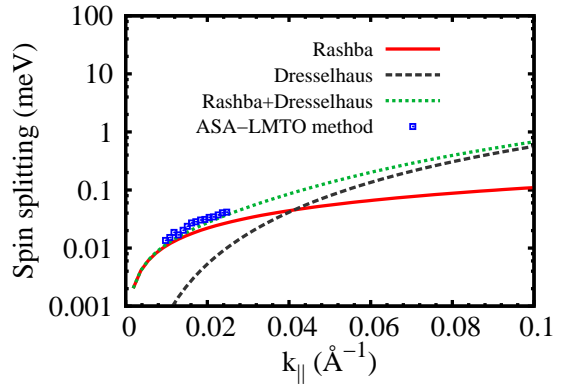}
 \caption{(Color Online) Electron spin splitting
    in ZnO against momentum along $\Gamma$-$K$ direction.
 The splitting from the  Rashba term and
from the  ASA-LMTO calculation are taken from Ref.\ \onlinecite{Voon}.
  }
  \label{fig2}
\end{figure}

In order to verify the accuracy of our computation, we compare our results
with the {\it ab initio} calculations based on atomic sphere
approximation linear-muffin-tin-orbital (ASA-LMTO) method, as shown
in Fig.\ \ref{fig2}.
In the figure, the Dresselhaus splitting  is from our computation and
the Rashba splitting is calculated by using the Rashba coefficient extracted from
ASA-LMTO calculations by Lew Yan Voon {\em et al.}.\cite{Voon}
From Fig.\ \ref{fig2},
one can see that the sum of
these two terms is in very good agreement with the ASA-LMTO results,
which confirms the soundness of our Dresselhaus spin splitting
calculations. This agreement also confirms
that the  Rashba  and Dresselhaus coefficients of electrons
have opposite signs.

\begin{figure}[bth]
  \centering
  \includegraphics[height=5.5cm]{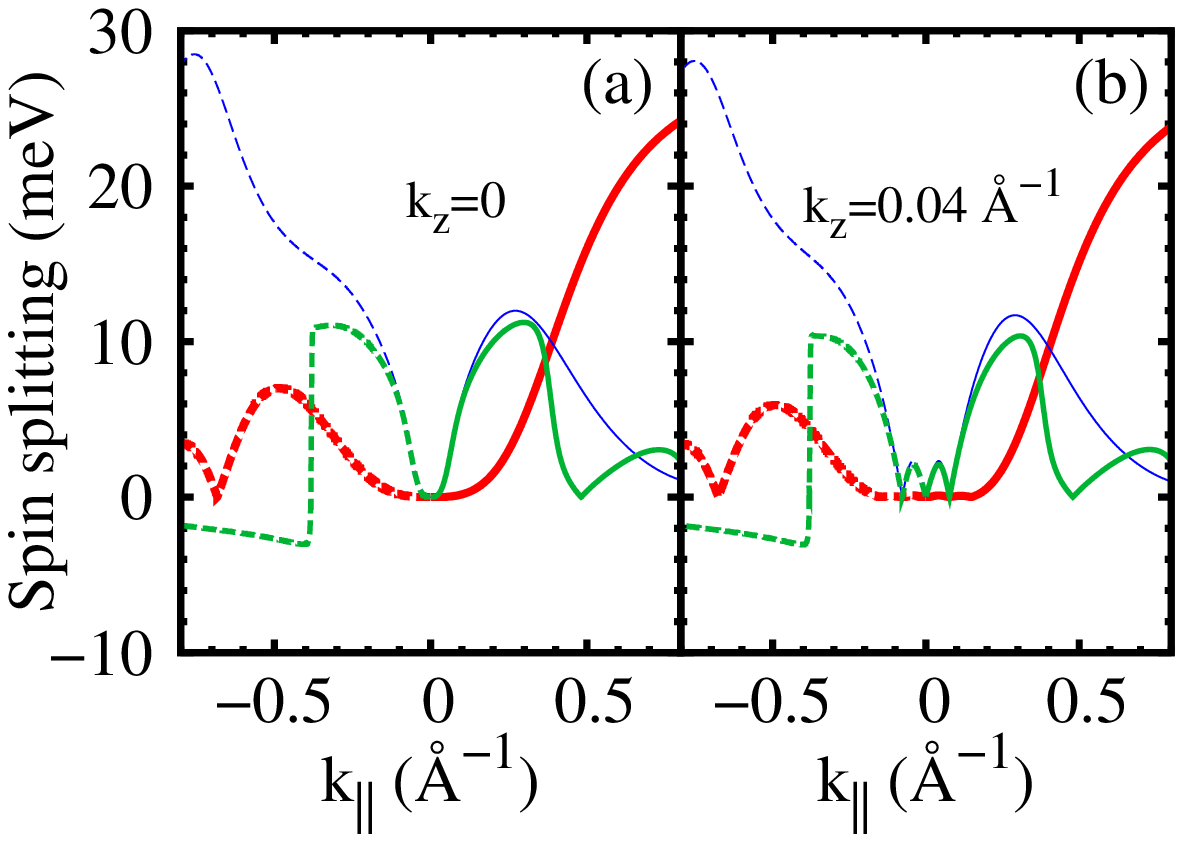}
  \includegraphics[height=5.cm]{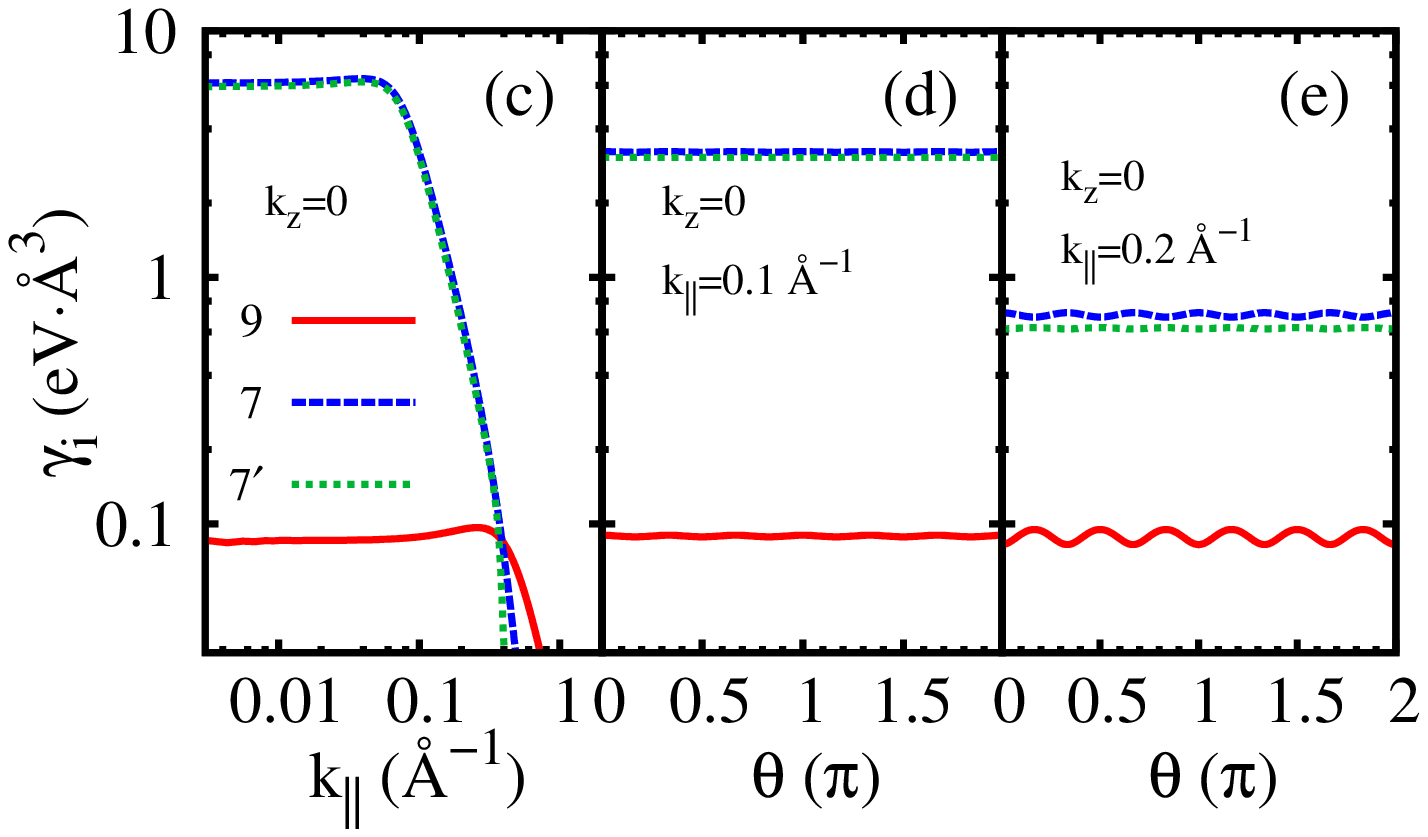}
 \caption{(Color Online) (a-b) Hole spin splitting in ZnO  against momentum
    along  $\Gamma$-$K$ (solid
    curves) and $\Gamma$-$M$ (dashed curves) directions with
 $k_z$=0 in (a) and  $k_z$=0.04\ \AA$^{-1}$ in (b).
(c) Hole SOC coefficients $\gamma_i$  ($i=9,7,7^\prime$)
{\it vs}. $k_{\|}$ along $\Gamma$-$K$ direction;
   (d-e) $\gamma_i$ {\it vs}. angle of the momentum in the plane of
   $k_z$=0. $k_{\|}$=0.1\ \AA$^{-1}$ in (d) and $k_{\|}$=0.2\ \AA$^{-1}$
in (e).  Red curves [thick curves in (a) and (b)]:  $\Gamma_{9v}$; Blue curves
[thin curves in (a) and (b)]:
$\Gamma_{7v}$; Green curves: $\Gamma_{7^\prime v}$.
 }
  \label{fig3}
\end{figure}

We now turn to the case of hole SOC.
The spin splittings for holes
with $\Gamma_{9v}$, $\Gamma_{7v}$, and $\Gamma_{7^\prime v}$ symmetries
are plotted against in-plane momentum in Fig.\ \ref{fig3}(a) and (b)
in the planes of $k_z=0$
and 0.04~\AA$^{-1}$ respectively. One
finds from Fig.\ \ref{fig3}(a) that
as $k_{\|}<0.3$\ {\AA}$^{-1}$, the spin splittings for the holes with
$\Gamma_{7v}$ and $\Gamma_{7^\prime v}$ symmetry
 increase drastically and become  much larger than that of
the holes with $\Gamma_{9v}$ symmetry.
 The corresponding SOC coefficients $\gamma_i$, with $i=9,7,7^\prime$
are also calculated similar to the case of $\gamma_e$.
The results are shown in
Fig.\ \ref{fig3}(c-e).
Again these coefficients are constants close to the
$\Gamma$ point, where $\gamma_7$ and $\gamma_{7^\prime}$ are close to each
other,  i.e., $\gamma_7=6.3$~eV$\cdot${\AA}$^{3}$
and $\gamma_{7^\prime}=6.1$~eV$\cdot${\AA}$^{3}$.
While for the hole with $\Gamma_{9v}$ symmetry, the coefficient
$\gamma_9$ is only about 0.09\ eV$\cdot${\AA}$^{3}$. Moreover,
 $\gamma_i$ are also angle-independent for small momentum,
but become markedly angle-dependent
for large momentum.
The corresponding results are shown in Fig.~\ref{fig3}(d) and (e).
Additionally, from the zero spin splitting point in
Fig.\ \ref{fig3}(b)
for the holes with  $\Gamma_{7v}$ and
$\Gamma_{7^\prime v}$ symmetries
(around 0.098\ {\AA}$^{-1}$ in the figure),
the parameter $b$ can also be extracted.
We obtain almost the identical value as that
obtained from the electron SOC under the same momentum.

\begin{figure}[bth]
 \includegraphics[height=5.5cm]{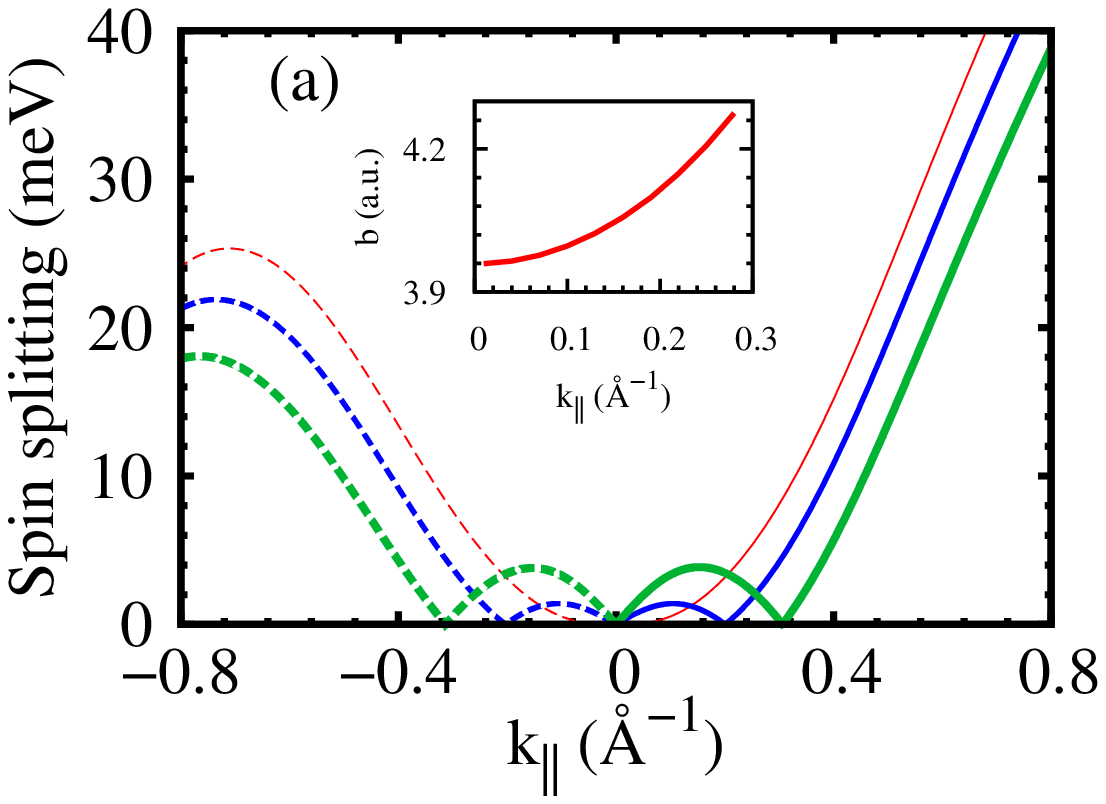}
 \includegraphics[height=5.5cm]{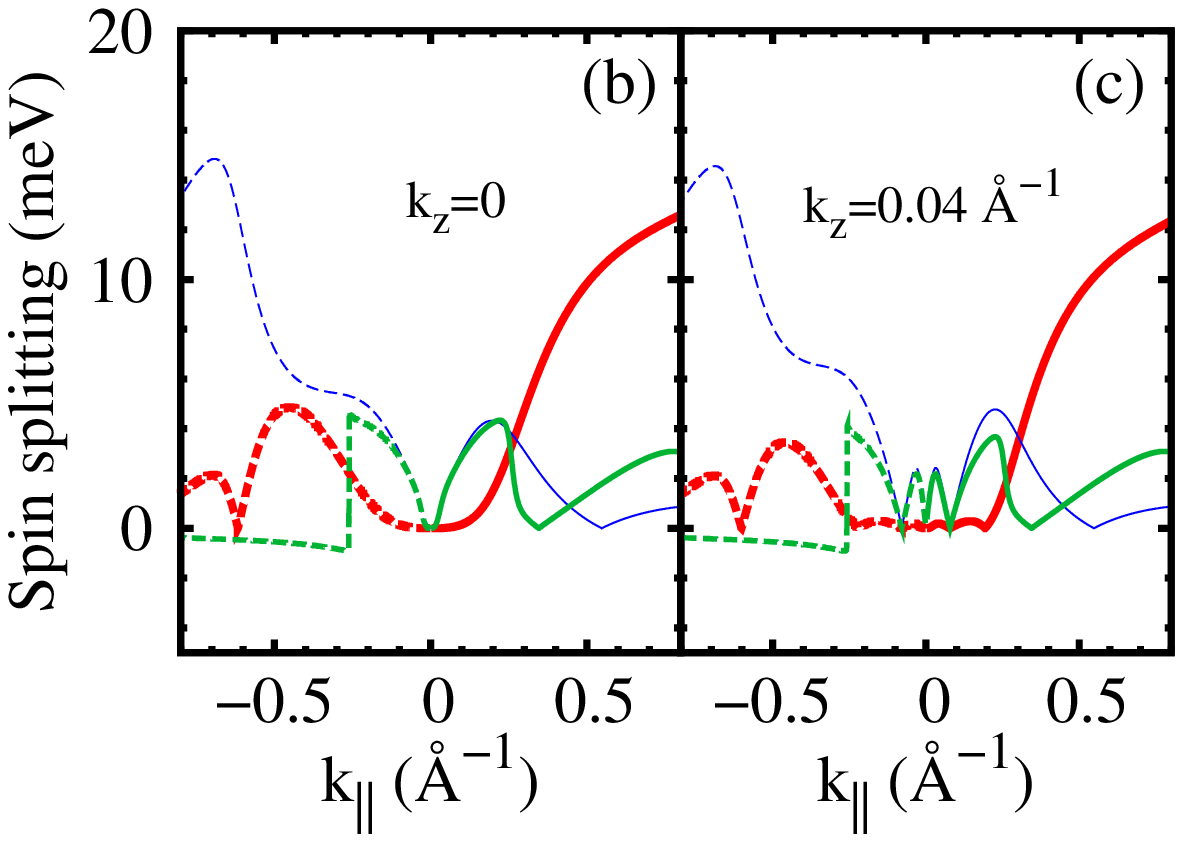}
 \caption{(Color Online) (a) Electron spin splitting in wurtzite
    GaN against momentum along  $\Gamma$-$K$ (solid
    curves) and $\Gamma$-$M$ (dashed curves)
    directions. Red (thin) curves: $k_z$=0; Blue curves: $k_z$=0.1\ {\AA}$^{-1}$; Green
  (thick)  curves: $k_z$=0.15\ {\AA}$^{-1}$.
Inset: The coefficient $b$ as function of
$k_\|$. (b-c) Hole spin splitting in wurtzite GaN  against momentum
    along  $\Gamma$-$K$ (solid
    curves) and $\Gamma$-$M$ (dashed curves) directions with
 $k_z$=0 in (b) and  $k_z$=0.04\ \AA$^{-1}$ in (c).
  Red (thick) curves:  $\Gamma_{9v}$; Blue (thin) curves:
$\Gamma_{7v}$; Green curves: $\Gamma_{7^\prime v}$.
  }
\label{fig4}
\end{figure}

\begin{table}[htbp]
 \caption{Rashba  ($\alpha_i$) (in meV$\cdot$\AA) and Dresselhaus ($\gamma_i$) (in
  eV$\cdot$\AA$^3$) SOC coefficients in wurtzite ZnO and GaN at small momentum
with $i=e,9,7,7^\prime$.}
\begin{tabular}{lllllllllllllllllll}\hline\hline
&\mbox{} &$\alpha_e$ & \mbox{}& $\alpha_9$ & \mbox{}&$\alpha_7$ & \mbox{}
& $\alpha_{7^\prime}$ & $\gamma_e$ &\mbox{}& $\gamma_9$\mbox{}& &$\gamma_7$\mbox{} &$\gamma_{7^\prime}$ \\ \hline
ZnO: &\mbox{} &1.1$^{\mbox{a}}$&\mbox{}& $-$ &\mbox{}& 35$^{\mbox{a}}$ (21$^{\mbox{b}}$) &\mbox{} &51$^{\mbox{a}}$ &0.33
 &\mbox{} &0.09&\mbox{}& 6.3 & 6.1 & \\
GaN: &\mbox{} &9.0$^{\mbox{b}}$&\mbox{}&$-$ &\mbox{}& 45$^{\mbox{b}}$ &\mbox{} & 32$^{\mbox{b}}$ &0.32
 &\mbox{} &0.07&\mbox{}& 15.3 & 15.0 & \\
\hline\hline\\
\end{tabular}
\label{table2}\\
$^{\mbox{a}}$ from Ref.\ \onlinecite{Voon}; \hspace{1cm}
$^{\mbox{b}}$ from Ref.\ \onlinecite{Majewski}
\end{table}

\subsection{SOC  in GaN}
GaN shares the same electronic wurtzite structure with ZnO, and also has a direct
wide gap about 3.4\ eV. \cite{Maruska}
In Fig.\ \ref{fig4}(a),
we show the electron Dresselhaus spin splittings
around $\Gamma$ point for different momentums
in the planes of $k_z$=0 (red thin curves), 0.1 (blue
curves), and 0.15\ \AA$^{-1}$ (green thick
curves), respectively. One can see from the
figure that the momentum dependence of the Dresselhaus splitting
is very similar to that in ZnO.
In GaN, the Rashba coefficient is
also small (about 9\ meV$\cdot${\AA}). We can also estimate that,
the Dresselhaus effect becomes  dominant when the
electron concentration is larger than  $10^{20}$~cm$^{-3}$.
 From the point of zero spin splitting,
one can calculate  $b$ in the Dresselhaus term.
The results are shown
in the inset of Fig.\ \ref{fig4}(a).
For small momentum, one finds $b\sim 3.959$ in GaN and 3.855 previously
in ZnO, both are different but close to 4.028---the ``universal''
value for all the wurtzite materials predicted by Wang {\em et al.} from their
AlN investigation.\cite{Wang} Moreover, with
the increase of the momentum, similar to the
case of ZnO,  $b$ can not be regarded as a constant and shows marked
$k$-dependence. In addition, it should be mentioned
that, for large momentum,
the universality of $b$ for all the wurtzite materials does not hold
and this difference enhances with the increase of momentum. For example,
when  $k_\|=0.3$\ \AA$^{-1}$, the values for ZnO and GaN are
 4.039 and 4.326, respectively.

\begin{figure}[bth]
  \includegraphics[height=5.5cm]{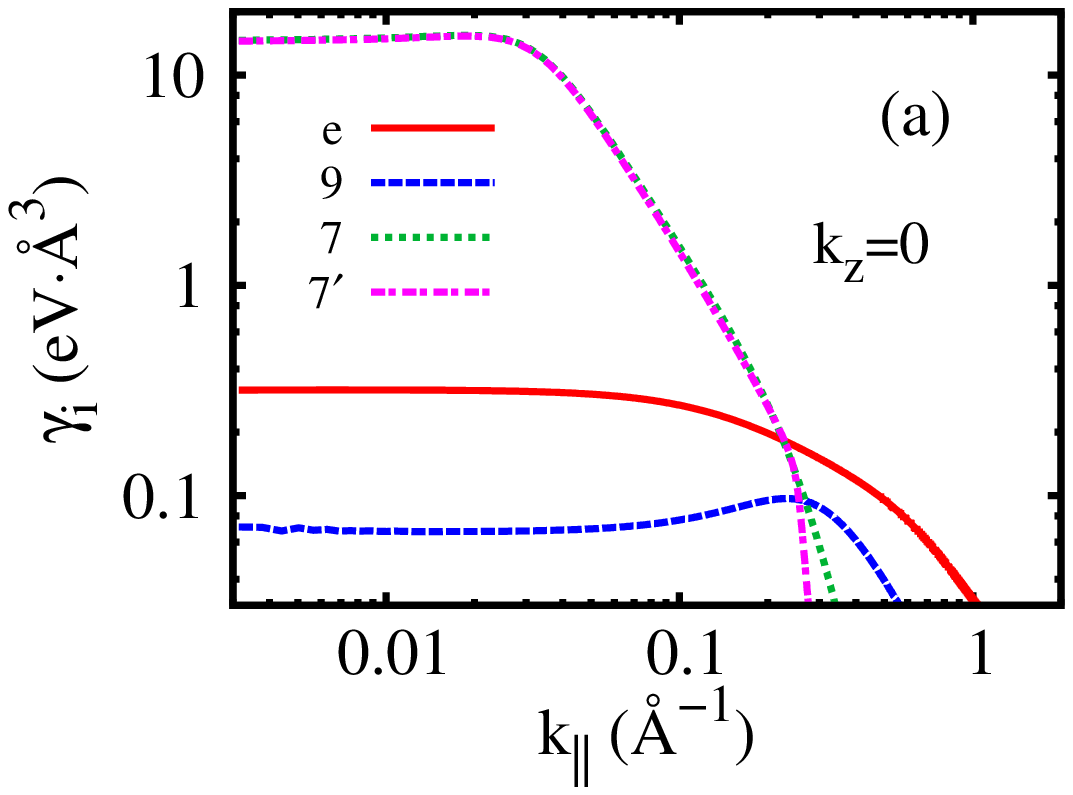}
  \includegraphics[height=5.5cm]{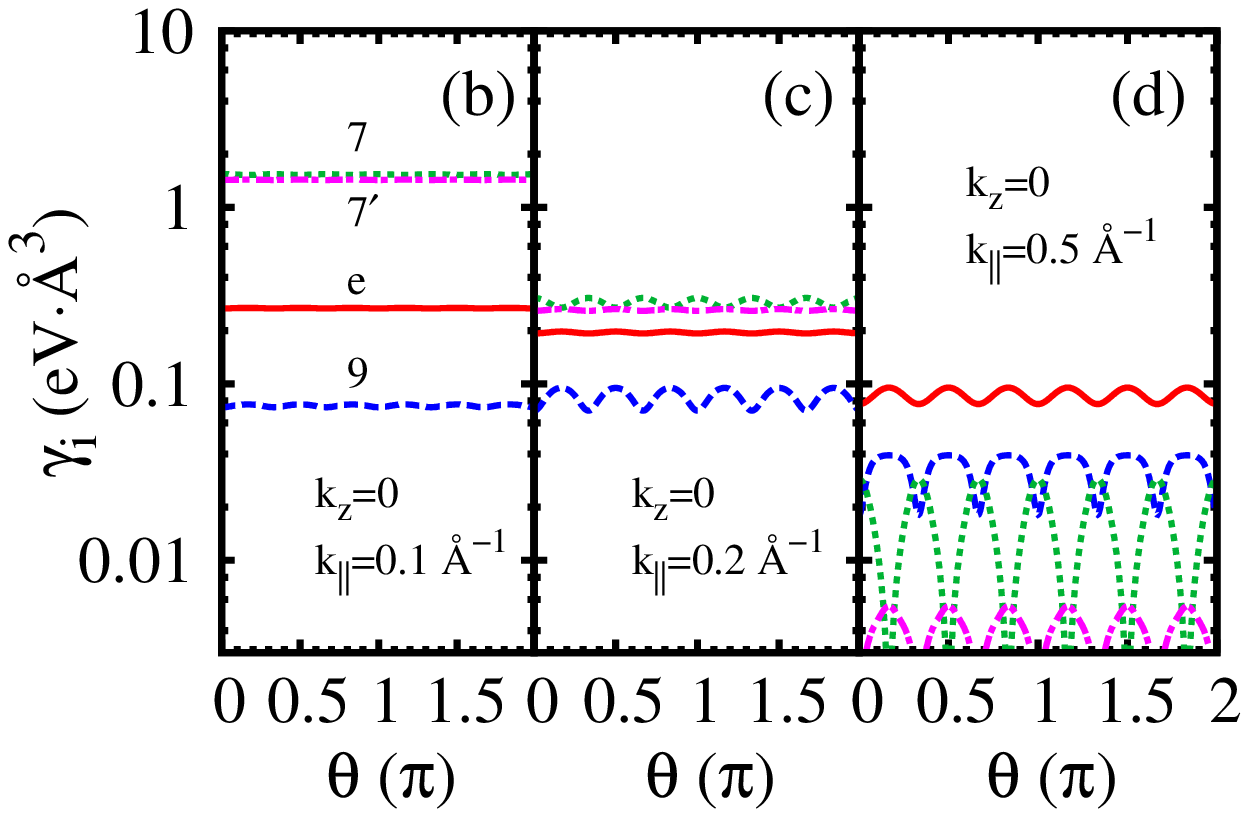}
 \caption{(Color Online)
(a) The SOC coefficients $\gamma_i$ with $i=e,9,7,7^\prime$
{\it vs}. $k_{\|}$ along $\Gamma$-$K$ direction with $k_z$=0;
    (b-d) $\gamma_i$ {\it vs}. angle of the momentum in the plane of
    $k_z$=0 with $k_{\|}$=0.1\ {\AA}$^{-1}$ (b), $k_{\|}$=0.2\ {\AA}$^{-1}$
(c), and $k_{\|}$=0.5\ {\AA}$^{-1}$ (d).
  }
  \label{fig5}
\end{figure}

\begin{figure}[bth]
  \centering
 \includegraphics[height=5cm]{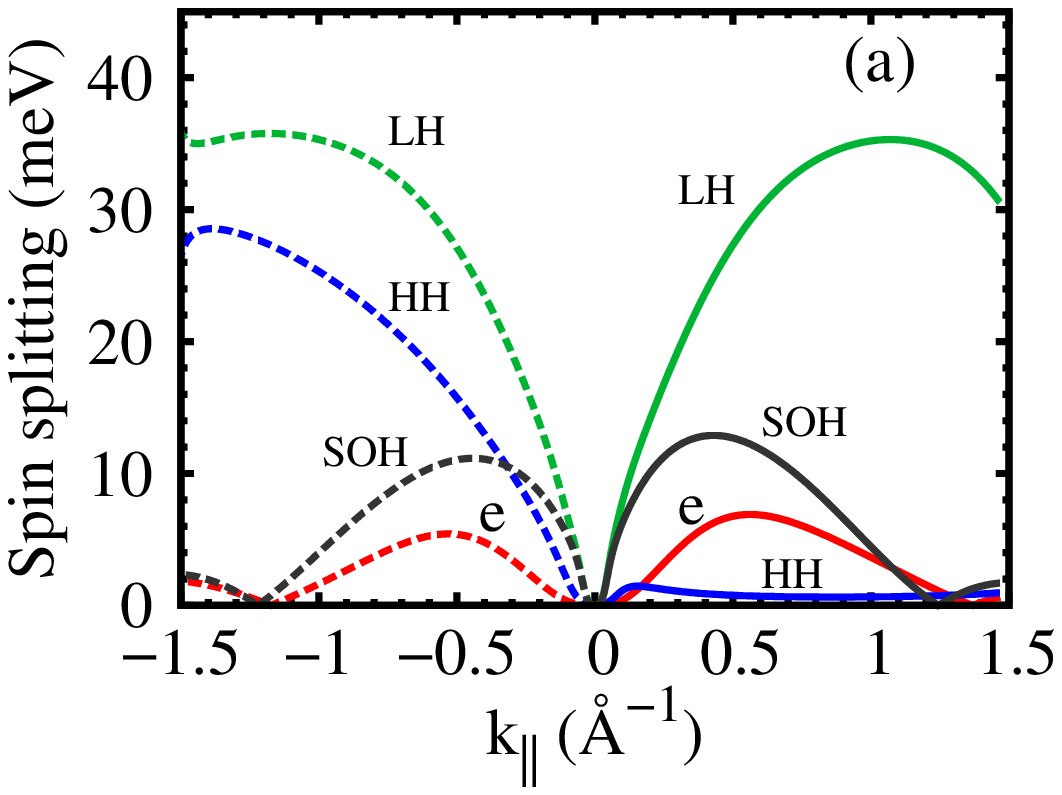}
 \includegraphics[height=5cm]{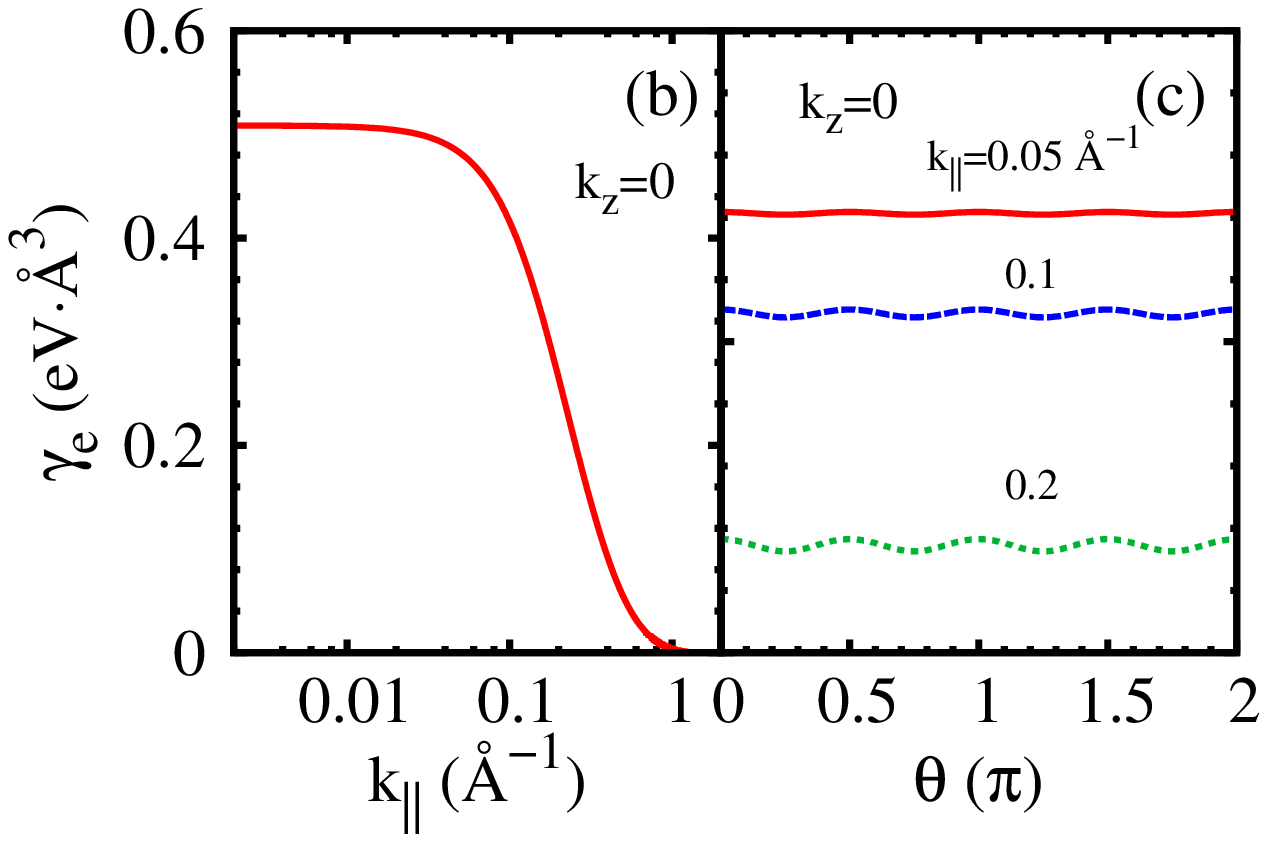}
 \includegraphics[height=5cm]{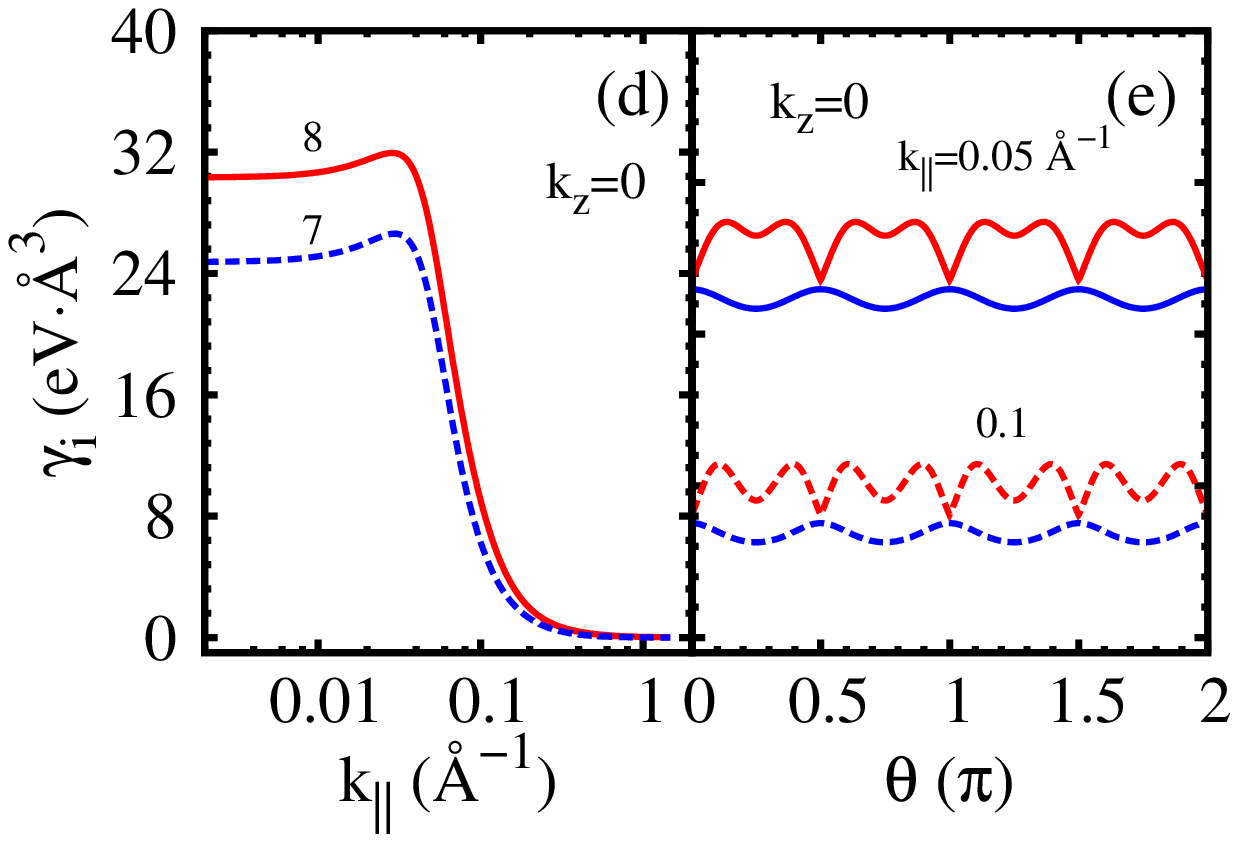}
\caption{(Color Online) (a) Electron and hole spin splitting in zinc-blende GaN
      against momentum
    along  $\Gamma$-$K$ (solid curves) and  
   $\Gamma$-$W$ (dashed curves) directions. Red curves:
     electron; Blue curves: HH; Green curves: LH; Black curves: SOH.
 (b) The corresponding electron SOC coefficient $\gamma_e$ {\it
vs}. $k_{\|}$ along  $\Gamma$-$K$ direction.
(c) $\gamma_e$ {\it vs}. the angle of the momentum at different $k_\|$
($k_z=0$).
    (d) $\gamma_8$ (red curve) and $\gamma_7$ (blue curve)
{\it vs}. momentum along  $\Gamma$-$K$ direction with $k_z=0$.
 (e) $\gamma_8$ (red curve) and $\gamma_7$ (blue curve) {\it vs}.
the angle of the momentum at different $k_\|$. Solid curves:
 $k_{\|}$=0.05; Dashed curves: $k_{\|}$=0.1\ {\AA}$^{-1}$.
  }
  \label{fig6}
\end{figure}

For the hole spin splitting, one also obtains the
similar results as those of ZnO, as shown in
Fig.\ \ref{fig4}(b) and (c). From Fig.\ \ref{fig4},
it is implied that both the electron and hole
spin splittings are isotropic at small momentum.
The corresponding SOC coefficients $\gamma_i$ with $i=e,9,7,7^\prime$
as functions of in-plane  momentum  are shown
in Fig.\ \ref{fig5}(a)
with $k_z=0$. We  find  $\gamma_e$ and $\gamma_9$ are about 0.32 and
0.07\ eV$\cdot${\AA}$^{3}$ for small momentum, respectively. These values
are very close to the case of ZnO, which we believe are due to their
similar electronic structures and almost equal band
gaps. However, as for $\gamma_7$ and $\gamma_{7^\prime}$, they are about 15.3 and
15.0\ eV$\cdot${\AA}$^{3}$ separately, two times larger than those
in ZnO. Similarly,  $\gamma_7$ and $\gamma_{7^\prime}$ are  close to
each other. In addition, as the momentum increases, the
dependence of the SOC coefficients on the
angle of the  in-plane-momentum
becomes more and more pronounced, as shown
in Fig.\ \ref{fig5}(b-d).
The Rashba ($\alpha_i$)  and Dresselhaus ($\gamma_i$)
SOC coefficients in wurtzite ZnO and GaN for small momentum  are summarized
in Table\ \ref{table2}.

In order to have a comparison with the case of zinc-blende
structures, we also calculate the spin splitting energy in cubic
GaN using $sp^3s^\ast$ TB model based on the parameters by Reilly {\em et
al.},\cite{Reilly} which has proven its capability in the
$\Gamma$ valley band structure calculations.\cite{Wu}
Both the electron and hole SOCs at the $\Gamma$ point in zinc-blende crystals
have been extensively discussed in Ref.\ \onlinecite{Winkler} and it is
known that only the Dresselhaus term exists in this structure.

The numerical results are shown in Fig.\ \ref{fig6}.
Two typical directions (along $\Gamma$-$K$ and  $\Gamma$-$W$)
of spin splittings for electron and hole are shown
in Fig.\ \ref{fig6}(a). It is implied that
even though for small momentum, the
spin splittings also show anisotropy.
This is not in the case of wurtzite structure.
Moreover, the anisotropic spin splitting is more distinct for the  HH,
as shown in Fig.\ \ref{fig6}(a).
Of course, for large momentum, the spin splittings of both structures show anisotropy.
In addition, it is interesting to see that
although the lowest conduction band and the SOH
band have different symmetries
($\Gamma_{6c}$ and $\Gamma_{7v}$ respectively), they
share the same form of SOC. From the red and black
curves in Fig.\ \ref{fig6}(a),
one can see that  electron and SOH have
similar momentum dependence of
spin splitting.

The electron SOC coefficient is shown in Fig.\ \ref{fig6}(b) and (c).
Similar to the wurtzite case,
$\gamma_e$ keeps  constant near the $\Gamma$ point and is independent
on the direction of the in-plane momentum. Here we obtain
$\gamma_e\sim0.51$\ eV$\cdot${\AA}$^{3}$, which is also
nearly two orders of magnitude smaller than that in GaAs.\cite{Wu}
The big difference comes from  the smaller
band gap of GaAs. When $k_{\|}>$ 0.03\ {\AA}$^{-1}$,
 $\gamma_e$ decreases drastically
with the momentum, and its angle dependence also becomes
very pronounced.

The corresponding hole SOC coefficients are
shown in Fig.\ \ref{fig6}(d) and (e).
Near the $\Gamma$
point, we find $\gamma_8$ and $\gamma_7$ around
30 and 25\ eV$\cdot${\AA}$^{3}$, respectively. Additionally, $\gamma_8$
shows stronger angle dependence of the momentum than
$\gamma_7$.

\section{Conclusion}

In conclusion, we have investigated
the electron and hole SOC
up to third order in wurtzite semiconductors.
Due to the intrinsic structure inversion asymmetry in addition to the bulk
inversion asymmetry in wurtzite structure,
both the linear Rashba effect and cubic Dresselhaus
effect exist. We find
the relative signs of the Rashba and Dresselhaus terms 
are opposite for electrons but are the same for  holes in both ZnO
and GaN. For  holes with $\Gamma_{9v}$ symmetry, there is no linear Rashba
contribution. Due to the negative spin splitting in ZnO, the signs of
the Rashba coefficients for both  electrons and holes
are opposite with those in GaN.

The Dresselhaus spin splittings
in wurtzite ZnO and GaN are calculated by the nearest-neighbor $sp^3$
TB model. We find that the electron Dresselhaus term
becomes the dominant SOC
compared with the linear
Rashba term when the electron concentration is higher than
$10^{19}$\ cm$^{-3}$ for ZnO and $10^{20}$\ cm$^{-3}$ for GaN.
While for the $\Gamma_{7v}$ and $\Gamma_{7^\prime v}$ holes, the linear term can
not be neglected due to the relatively larger hole Rashba SOC coefficient.
The corresponding SOC coefficients $\gamma_i$ are also calculated, which
are momentum-independent  very near the $\Gamma$ point.
However, at large momentum, they decrease drastically  and show strong
angular  dependence of the in-plane momentum.
In addition, we confirm that the Dresselhaus parameter $b$ is
almost a constant for very small momentum and  can be regarded approximately
as an universal
value for all the wurtzite materials as predicted in Ref.\
\onlinecite{Wang}.
However, for larger momentum, its $k$-dependence can not be neglected
and the value varies strongly for different materials.

In order to compare with the zinc-blende structure,
the spin splittings in cubic GaN are also addressed,
calculated using $sp^3s^\ast$ model. The isotropic
spin splitting near the $\Gamma$ point in
wurzite structures does not hold in zinc-blende structure.

We believe that our results are useful for the on-going study
of spin dynamics of ZnO and GaN.
After all, due to the small SOC in these wide gap semiconductors,
the spin relaxation time becomes very long and hence more potential applications
may be figured out using these semiconductors.
Finally, we point out that due to the lack of the TB parameters for
ZnO of zinc-blende structure, we cannot calculate the SOC coefficients
for this structure. More investigations are need.

\begin{acknowledgments}
This work was supported by the Natural
Science Foundation of China under Grants No.\ 10574120 and
No.\ 10725417, the
National Basic Research Program of China under Grant
No.\ 2006CB922005, the Knowledge Innovation Project of Chinese Academy
of Sciences, and the 
Robert-Bosch Stiftung. One of the authors (M.W.W.) would like to thank
X. Marie and U. Schwarz
for bringing this topic into his attention as
well as valuable discussions.
J.Y.F. was also  partially supported by China Postdoctoral
Science Foundation.
\end{acknowledgments}


\begin{thebibliography}{0}
\bibitem{Bagnall} D. M. Bagnall, Y. F. Chen, Z. Zhu, T. Yao,
  S. Koyama, M. Y. Shen, and T. Goto,
    Appl. Phys. Lett. {\bf 70}, 2230 (1997).
\bibitem{Jain} S. C. Jain, M. Willander, J. Narayan, and R. Van Overstraeten,
    J. Appl. Phys. {\bf 87}, 965 (2000).
\bibitem{Vurgaftman} I. Vurgaftman and J. R. Meyer,
    J. Appl. Phys. {\bf 94}, 3675 (2003).
\bibitem{rev1}A. Ashrafi and C. Jagadish, J. Appl. Phys. {\bf 102}, 071101
(2007); C. Klingshirn, phys. stat. sol. (b) {\bf 244}, 3027 (2007).
\bibitem{Shen1} K. S. Cho, Y. F. Chen, Y. Q. Tang, and B. Shen,
   Appl. Phys. Lett. {\bf 90}, 041909 (2007).
\bibitem{Shen2} X. W. He, B. Shen, Y. Q. Tang, N. Tang, C. M. Yin,
  F. J. Xu, Z. J. Yang, G. Y. Zhang, Y. H. Chen, C. G. Tang, and
  Z. G. Wang,
   Appl. Phys. Lett. {\bf 91}, 071912 (2007).

\bibitem{Yan} S. S. Yan, C. Ren, X. Wang, Y. Xin, Z. X. Zhou,
  L. M. Mei, M. J. Ren, Y. X. Chen, Y. H. Liu, and H. Garmestani,
    Appl. Phys. Lett. {\bf 84}, 2376 (2004).
\bibitem{Ghosh} S. Ghosh, V. Sih, W. H. Lau, D. D. Awschalom,
  S.-Y. Bae, S. Wang, S. Vaidya, and G. Chapline,
    Appl. Phys. Lett. {\bf 86}, 232507 (2005).
\bibitem{Liu} W. K. Liu, K. M. Whitaker, A. L. Smith,
  K. R. Kittilstved, B. H. Robinson, and D. R. Gamelin,
    Phys. Rev. Lett. {\bf 98}, 186804 (2007).
\bibitem{Dietl} T. Dietl, H. Ohno, F. Matsukura, J. Cibert, and D. Ferrand,
    Science {\bf 287}, 1019 (2000).

\bibitem{Beschoten} B. Beschoten, E. Johnston-Halperin, D. K. Young,
  M. Poggio, J. E. Grimaldi, S. Keller, S. P. DenBaars, U. K. Mishra,
  E. L. Hu, and D. D. Awschalom,
    Phys. Rev. B {\bf 63}, 121202(R) (2001).
\bibitem{Lagarde} D. Lagarde, A. Balocchi, P. Renucci, H. Carr\`ere,
  F. Zhao, T. Amand, X. Marie, Z. X. Mei, X. L. Du, and Q. K. Xue,
    arXiv:0804.2369.
\bibitem{Wolf} S. A. Wolf, J. Supercond. {\bf 13}, 195 (2000).

\bibitem{Zutic} I. \v{Z}uti\'c, J. Fabian, S. D. Sarma,
     Rev. Mod. Phys. {\bf 76}, 323 (2004); J. Fabian, A. Matos-Abiague, C. Ertler,
P. Stano, and I. \v Zuti\'c, Acta Physica Slovaca {\bf 57}, 565 (2007).
\bibitem{Bir} G. L. Bir and G. E. Pikus, {\em Symmetry and
Strain-Induced Effects in Semiconductors} (Wiley, New York, 1974).
\bibitem{Weber} W. Weber, S. D. Ganichev, S. N. Danilov, D. Weiss,
  W. Prettl, Z. D. Kvon, V. V. Bel'kov, L. E. Golub, Hyun-lck Cho, and
  Jung-Hee Lee,   Appl. Phys. Lett. {\bf 87}, 262106 (2005).
\bibitem{Rashba} E. I. Rashba,
     Fiz. Tverd. Tela (Leningrad) {\bf 2}, 1224 (1960) [Sov. Phys.
     Solid State {\bf 2}, 1190 (1960)].
\bibitem{Bychkov}Y. A. Bychkov and E. I. Rashba, Pis'ma
     Zh. Eksp. Teor. Fiz. {\bf 39}, 66 (1984) [Sov. Phys. JETP
     Lett. {\bf 39},  78  (1984)].
\bibitem{Voon} L. C. Lew Yan Voon, M. Willatzen, and M. Cardona,
   N. E. Christensen,   Phys. Rev. B {\bf 53}, 10703 (1996).
\bibitem{Majewski} J. A. Majewski and P. Vogl,
      {\em Physics of Semiconductors: $27^{th}$ International Conference on
        the Physics of Semiconductors}, edited by J. Men\'endez and
      C. G. Van de Walle (American Institute of Physics, 2005),
      p.1403.
\bibitem{Kurdak}C. Kurdak, N. Biyikli, \"U. \"Ozg\"ur, H. Morkoc and
    V. I. Litvinov,
     Phys. Rev. B {\bf 74}, 113308 (2006).
\bibitem{Thillosen} N. Thillosen, Th. Sch\"apers, N. Kaluza,
  H. Hardtdegen, and V. A. Guzenko,
     Appl. Phys. Lett. {\bf 88}, 022111 (2006).
\bibitem{Schmult} S. Schmult, M. J. Manfra, A. Punnoose,
  A. M. Sergent, K. W. Baldwin, and R. J. Molnar,
     Phys. Rev. B {\bf 74}, 033302 (2006).
\bibitem{Belyaev} A. E. Belyaev, V. G. Raicheva, A. M. Kurakin,
   N. Klein, and S. A. Vitusevich,
     Phys. Rev. B {\bf 77}, 035311 (2008).
 \bibitem{Wang} W. T. Wang, C. L. Wu, S. F. Tsay, M. H. Gau, I.
   Lo, H. F. Kao, D. J. Jang, J. C. Chiang, M. E. Lee, Y. C. Chang,
   C. N. Chen, and H. C. Hsueh,
    Appl. Phys. Lett. {\bf 91}, 082110 (2007).
\bibitem{Lowdin} P. L\"owdin,
      J. Phys. Chem. {\bf 19}, 1396 (1951).
\bibitem{Kobayashi} A. Kobayashi, O. F. Sankey, S. M. Volz, and
   J. D. Dow,  Phys. Rev. B {\bf 28}, 935 (1983).
\bibitem{Chadi} D. J. Chadi,
      Phys. Rev. B {\bf 16}, 790 (1977).
\bibitem{Chuang} S. L. Chuang and C. S. Chang, Phys. Rev. B {\bf 54}, 2491 (1996).
\bibitem{Kane} E. Q. Kane, J. Phys. Chem. Solids {\bf 1}, 249 (1957).
\bibitem{Winkler} R. Winkler, {\em Spin-Orbit Coupling Effects in
    Two-Dimensional Electron and Hole Systems} (Springer, Berlin, 2003).
\bibitem{Dugdale} D. J. Dugdale, S. Brand, and R. A. Abram,
       Phys. Rev. B {\bf 61}, 12933 (2000).
\bibitem{Ikai} I. Lo, W. T. Wang, M. H. Gau, S. F. Tsay, and
  J. C. Chiang,  Phys. Rev. B {\bf 72}, 245329 (2005).
\bibitem{Pfeffer} P. Pfeffer and W. Zawadzki,
      Phys. Rev. B {\bf 72}, 035325 (2005).
\bibitem{Zunger} S. H. Wei and A. Zunger,
      Appl. Phys. Lett. {\bf 69}, 2719 (1996).
\bibitem{Kumagai} M. Kumagai, S. L. Chuang, and H. Ando,
      Phys. Rev. B {\bf 57}, 15303 (1998).
\bibitem{Xia} W. J. Fan, J. B. Xia, P. A. Agus, S. T. Tan, S. F. Yu,
  and X. W. Sun, J. Appl. Phys. {\bf 99}, 013702 (2006).
\bibitem{Rowe} J. E. Rowe, M. Cardona, and F. H. Pollak,
      Solid State Commun. {\bf 6}, 239 (1968).
\bibitem{Casella} R. C. Casella,
      Phys. Rev. Lett. {\bf 5}, 371 (1960).
\bibitem{Mahan} G. D. Mahan and J. J. Hopfield,
      Phys. Rev. {\bf 135}, A428 (1964).
\bibitem{Jenkins} D. W. Jenkins and J. D. Dow,
      Phys. Rev. B {\bf 39}, 3317 (1989).
\bibitem{Malkova} N. Malkova and C. Z. Ning,
      Phys. Rev. B {\bf 75}, 155407 (2007).
\bibitem{Shindo} K. Shindo, A. Morita, and H. Kamimura,
      J. Phys. Soc. Jap. {\bf 20}, 2054 (1965).
\bibitem{Phillips} J. C. Phillips, {\em Bonds and Bands in Semiconductors},
    (Academic, San Diego, CA, 1973).
\bibitem{Izyumskaya} N. Izyumskaya, V. Avrutin, \"U. O\"zg\"ur,
  Y. I. Alivov, and H. Morkoc,
phys. stat. sol. (b) {\bf 244}, 1439 (2007).
\bibitem{Wu} J. Y. Fu, M. Q. Weng, and M. W. Wu,
      Physica E {\bf 40}, 2890 (2008).
\bibitem{Morkoc} H. Morkoc, {\em Nitride Semiconductors and Devices}
     (2nd ed., Springer, Berlin, 2007).
\bibitem{Maruska} H. P. Maruska and J. J. Tietjen,
      Appl. Phys. Lett. {\bf 15}, 327 (1969).
\bibitem{Reilly} E. P. O'Reilly, A. Lindsay, S. Tomi\'c, and
  M. K. Saadi,   Semicond. Sci. Technol. {\bf 17}, 870 (2002).

\end{thebibliography}
\end{document}